\documentclass {aa} 

\usepackage{amsmath}
\usepackage{graphicx}
\usepackage[title]{appendix}
\usepackage{float}
\usepackage[varg]{txfonts}
\usepackage{siunitx}
\usepackage{lscape}
\usepackage{bm}
\usepackage{rotating}
\usepackage{array}
\usepackage{natbib}
\usepackage{hyperref}
\hypersetup{
colorlinks=true,
linkcolor={blue},
citecolor={blue},
anchorcolor={blue},
urlcolor={blue}
}
\usepackage{siunitx}
\usepackage{xcolor}
\definecolor{Black}{rgb}{0.0,0.0,0.0}
\begin{document}

\title{Analyses of celestial pole offsets with VLBI, LLR, and optical observations}
\subtitle{}
\author{Y.-T. Cheng,\,
 J.-C. Liu,\,
 \and
 Z. Zhu}
\institute{School of Astronomy and Space Science, Key Laboratory of Modern Astronomy and Astrophysics (Ministry of Education), Nanjing University, 163 Xianlin Avenue, 210023 Nanjing, China\\
 \email{jcliu@nju.edu.cn}}

\date{}


\abstract
{}
{This work aims to explore the possibilities of  determining \textcolor{Black}{the long-period part of the precession-nutation} of the Earth with techniques other than very long baseline interferometry (VLBI). Lunar laser ranging (LLR) is chosen for its relatively high accuracy and long period. \textcolor{Black}{Results of previous studies could be updated using the latest data with generally higher quality, which would also add ten years to the total time span. Historical optical data  are also analyzed for their rather long time-coverage to determine whether it is possible to improve the current Earth precession-nutation model.}}
{Celestial pole offsets (CPO) series were obtained from LLR and optical observations and were analyzed separately by weighted least-square fits of three empirical models, including a quadratic model, a linear term plus an 18.6-year nutation term, and a linear term plus two nutation terms with 18.6-year and 9.3-year periods. \textcolor{Black}{Joint analyses} \textcolor{Black}{of VLBI and LLR data} is also presented for further discussion.}
{\textcolor{Black}{We improved th determination of the nutation terms with both VLBI and LLR data.} The VLBI data present \textcolor{Black}{a \textcolor{Black}{most reliable} feature of the CPO series with the highest accuracy,} and they are most important for determining the precession-nutation of the Earth. The standard errors of CPO obtained from the LLR technique have reached a level of several tens of microarcseconds after 2007, but \textcolor{Black}{they are probably} underestimated because the models used in the calculation procedure are not perfect. \textcolor{Black}{Nevertheless, the poor time resolution of LLR CPO series is also a disadvantage.} However, this indicates that LLR has \textcolor{Black}{the} potential \textcolor{Black}{to determine} celestial pole offsets with \textcolor{Black}{a comparably high accuracy with VLBI} in the future and to serve as an independent check for the VLBI results. \textcolor{Black}{The current situation of LLR observations is also analyzed to provide suggestions of future improvement.} The typical standard error of CPO series from historic optical observations is \textcolor{Black}{about two hundred times larger than that of the VLBI series and can therefore hardly contribute to the contemporary precession-nutation theory.}}
{}

\keywords{astrometry -- reference systems -- Moon -- \textcolor{Black}{VLBI}}

\maketitle
\section{Introduction}
\textcolor{Black}{Earth precession-nutation models describe the long-term and long-period changes of the celestial intermediate pole (CIP) direction in the \textcolor{Black}{geocentric celestial reference system (GCRS)}. Before 2000, the IAU\,1976/1980 precession-nutation model ~\citep{Lieske1977,Wahr1981,Seidelmann1982} was adopted, which is based on a celestial reference system based on the bright star catalog FK5. With the help of very long baseline interferometry (VLBI), these models were developed and thus finally replaced previous models with the IAU\,2006 precession~\citep{Capitaine2003} and IAU\,2000A nutation~\citep{Herring2002,Mathews2002} models.}

\textcolor{Black}{The CIP axis is used as a rotation axis to orient the international terrestrial reference system} with respect to a kinematically nonrotating celestial coordinate system, the latter being given by a group of distant radio source positions~\citep{Fey2015}. \textcolor{Black}{In this work, we focus on one part of the Earth orientation parameters (EOP), namely \textcolor{Black}{the celestial pole offsets (CPO), which represent} the differences between observations and \textcolor{Black}{theoretical predictions} of the CIP locations (i.e., d$X$, d$Y$).}

\textcolor{Black}{By far, VLBI has been playing the most crucial role in this domain, given its microarcsecond-} level accuracy. However,  VLBI observations have been available for a short time so far (about 38 years\textcolor{Black}{, 1979-2017}), which can be a disadvantage in revealing effects with long periods such as the deficiencies in the Earth precession model, which has a period of about \textcolor{Black}{26 000} years.

The satellite space-geodetic techniques, such as the Global Positioning System (GPS), lunar laser ranging (LLR), or satellite laser ranging (SLR), can also \textcolor{Black}{theoretically} define an inertial reference frame in a dynamical sense. LLR, which has been operated since 1969, is the only space-geodetic technique now capable to offer a stable dynamical reference system, with sufficient  determination accuracy of the lunar orbit (\citealt{Zerhouni2009}, denoted as ZC09 hereafter).
\textcolor{Black}{ZC09 provided a basic method} of determining CPO from LLR observations using data in the interval of 1969-2008. They have also estimated the corrections to nutation terms, compared the results with those of VLBI observations, \textcolor{Black}{and} presented \textcolor{Black}{joint analyses} of VLBI and LLR. The typical standard deviation of the fitted coefficients is about 0.1-0.2 mas for the nutation terms, and \textcolor{Black}{3 mas cy$^{-1}$ for the secular terms}. \citet{Hofmann2018} (denoted as H18 hereafter) \textcolor{Black}{have} estimated the reflector coordinates, station coordinates, and velocities and EOP \textcolor{Black}{at the same time} with LLR data during 1969-2016. The typical standard deviation for corrections to nutation terms in H18 is about $0.1$ mas. \textcolor{Black}{It was concluded in both works that the LLR technique lacks accuracy as well as sufficient observations to determine the Earth-related parameters with the same level of accuracy as VLBI.}

The history of optical observations, traced back to the nineteenth century, is much longer than that of either VLBI or LLR observations.~\citet{Vondrak2010} have constructed a series of Earth orientation catalogs (EOC) based on a combination of Hipparcos and Tycho catalogs and long lasting ground-based astrometric observations. With robust determinations of the proper motions of a binary star system from 4.5 million individual observations, the latest version, the EOC-4 catalog~\citep{Vondrak}, contains 4418 celestial objects (including stars, double star components, and photocenters). These catalogs were used to derive the \textcolor{Black}{CPO} during the time interval 1899.7-1992.0, in the Hipparcos celestial reference frame (HCRF), which can be regarded as the optical counterpart of the ICRS. The current EOP solution, denoted OA10~\citep{Vondrak}, was composed of a series of polar motion, UT1, \textcolor{Black}{but the CPO were only presented as quadratic functions of time. Thus, we used the CPO series of the penultimate version (OA00) in this work.}

\begin{figure*}
 \centering
 \includegraphics[width=0.8\linewidth]{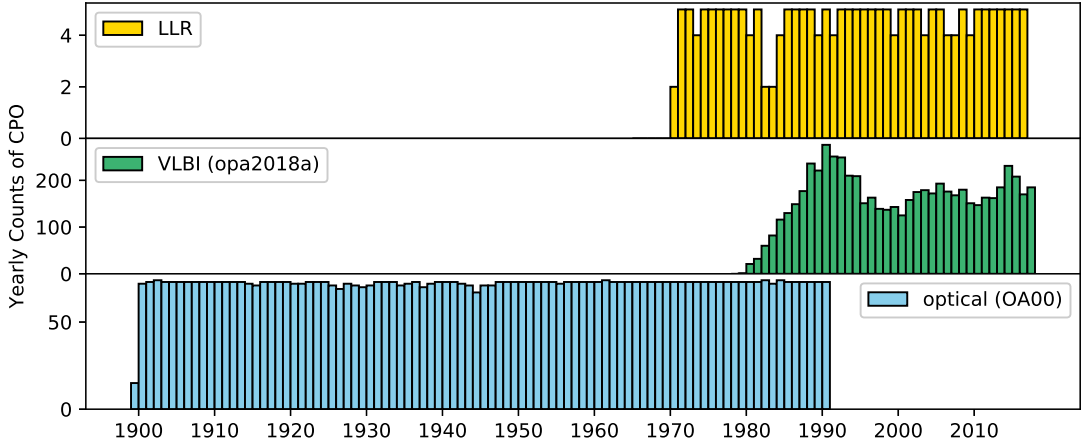}
 \caption{Time distributions of LLR, VLBI (opa2018a), and optical CPO series. \textcolor{Black}{The time span of every CPO series is LLR 1970.6-2017.7, VLBI (opa2018a) 1979.9-2018.5, and optical (OA00) 1899.7-1991.9.}}
 \label{fig-distribution}
\end{figure*}

We here \textcolor{Black}{aim} to improve the determination of precession-nutation corrections \textcolor{Black}{based on analyses of VLBI (Sect.~\ref{section_VLBI}), LLR (Sect.~\ref{section_LLR}), and optical (Sect.~\ref{section_optical}) observations, and we also present \textcolor{Black}{joint analyses} of VLBI and LLR data (Sect.~\ref{section_joint}). We follow the basic method of ZC09 for the analyses of LLR and the joint \textcolor{Black}{analyses} of LLR and VLBI, and make a comparison with the results of H18.} \textcolor{Black}{Further discussions and the potential of LLR} are presented in Sect.~\ref{section_discuss}. \textcolor{Black}{The time distribution of CPO series derived from LLR, VLBI (opa2018a), and optical observations are plotted in Fig.~\ref{fig-distribution} to show the difference\textcolor{Black}{s} in \textcolor{Black}{the number of observations} and in the time span.}
\section{Models used in the CPO analyses}
\label{section_models}
We considered several models that are commonly used in CPO analyses (e.g.,~\citealt{Capitaine2009}) to estimate corrections to the long-period \textcolor{Black}{part of the} precession-nutation models. All models are functions of $t$ (centuries from the basic epoch). The basic epoch was set to be J2000.0 for the VLBI and LLR series but to \textcolor{Black}{be} J1956.0 for optical series, \textcolor{Black}{which is almost the central epoch of the time coverage} of the optical observations.

\textcolor{Black}{Three empirical models were fit to the CPO series} in our \textcolor{Black}{analyses}:
\begin{enumerate}
 \item A parabola, that is, a quadratic function of $t$.
 \item \textcolor{Black}{A linear term} and 18.6-year nutation term:
       \begin{equation}
        \mathrm{d}X,\mathrm{d}Y = A_{\rm{0}} + A_{\rm{1}}t + A_{\rm{s}}\sin\Omega_1 + A_{\rm{c}}\cos\Omega_1,
       \end{equation}
 \item \textcolor{Black}{A linear term}, and 18.6-year and 9.3-year nutation terms:
       \begin{equation}
        \begin{split}
         \mathrm{d}X,\mathrm{d}Y = A_0 + A_1 t + A_{\rm{s1}}\sin\Omega_1 + A_{\rm{c1}}\cos\Omega_1 \\
         + A_{\rm{s2}}\sin\Omega_2 + A_{\rm{c2}}\cos\Omega_2,
        \end{split}
       \end{equation}
       where $\Omega_1 = 2\pi/0.186$ and $\Omega_2 = 2\pi/0.093$\textcolor{Black}{, representing the arguments of the two principle nutation terms}. \textcolor{Black}{The 18.6-year term is related to the motion of the lunar node.}
\end{enumerate}
All models were used in VLBI analyses for comparison with the other two techniques. Models 2 and 3 are used in LLR analyses to estimate corrections to the IAU\,2006/2000 precession-nutation models. Models 1 and 2 are used in optical analyses to estimate effects with relatively longer periods.

\section{CPO analyses with VLBI observations}
\label{section_VLBI}
\begin{figure}[!htbp]
 \centering
 \includegraphics[width=8cm]{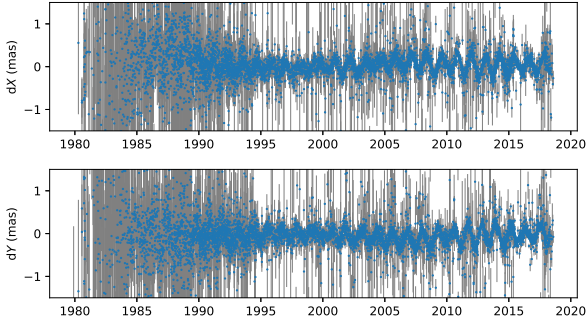}
 \caption{VLBI celestial pole offsets from the OPA analysis center with respect to the IAU\,2006/2000 models.}
 \label{OPA}
\end{figure}

\begin{figure}[!htbp]
 \centering
 \includegraphics[width=8cm]{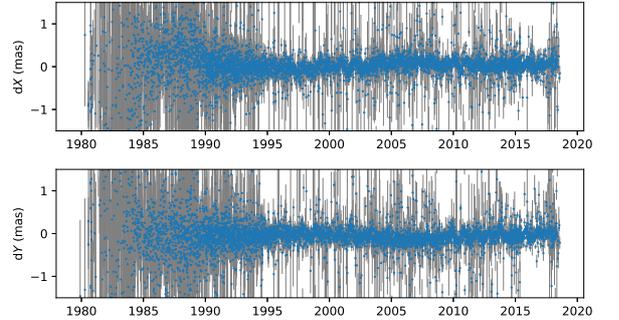}
 \caption{d$X$ and d$Y$ residuals of the OPA VLBI time series with respect to the IAU\,2006/2000 models. FCN is removed in the updated IERS model.}
 \label{OPA-FCN}
\end{figure}

We used the CPO series from 1979 to 2018 derived by the Calc/Solve software at the Paris Observatory analysis center (OPA)\footnote{\url{http://ivsopar.obspm.fr/24h/opa2018a.eops}} (Fig.~\ref{OPA}) and also the series derived by the Goddard Space Flight Center (GSF)\footnote{\url{https://vlbi.gsfc.nasa.gov/solutions/itrf/2016a/gsf2016a.eops}} (not plotted because it is very similar to the OPA series) in the framework of the international VLBI service for geodesy and astronomy (IVS, \citet{IVS}). The quasi-periodic free core nutation (FCN), \textcolor{Black}{with a period of 430.23 days,} stands out in the structures of the series, and needs to be removed before the residuals are analyzed. We removed this effect according to the model recommended by the IERS conventions 2010 \citep{Petit2010} (updated yearly\footnote{\url{http://ivsopar.obspm.fr/24h/geo.txt}}), and the OPA series without FCN are shown in Fig.~\ref{OPA-FCN}.
\begin{table*}[!htbp]
 \caption{Weighted fits of three models to VLBI residuals (1979.6-2018.6), corresponding to the IAU\,2006/2000 model.}
 \label{tab_VLBI}
 \centering
 \scalebox{0.9}{
  \begin{tabular}{rrrrrrrrrrr}
   \hline\hline
            \textcolor{Black}{Term}&      & $t^0$                         & $t^1$                          & $t^2$                           & sin(18.6 yr)                 & cos(18.6 yr)                 & sin(9.3 yr)                  & cos(9.3 yr)                 & WRMS$_{\rm pre}$        & WRMS$_{\rm post}$       \\
            \textcolor{Black}{Unit}&      & \si{\mu as}                   & \si{\mu as}/cy                 & \si{\mu as}/cy$^2$                & \si{\mu as}                  & \si{\mu as}                  & \si{\mu as}                  & \si{\mu as}                 & \si{\mu as}              & \si{\mu as}              \\
   \hline
   opa2018a & d$X$ & $18\pm1$                      & $266\pm21$                     & $-4338\pm143$                   &                              &                              &                              &                             & $126$                    & $119$                    \\
            & d$Y$ & $-86\pm1$                     & $-519\pm22$                    & $5325\pm145$                    &                              &                              &                              &                             & $147$                    & $130$                    \\
   \hline

   gsf2016a & d$X$ & \textcolor{Black}{$49\pm1$}   & \textcolor{Black}{$215\pm19$}  & \textcolor{Black}{$540\pm120$}  &                              &                              &                              &                             & \textcolor{Black}{$134$} & \textcolor{Black}{$104$} \\
            & d$Y$ & \textcolor{Black}{$-101\pm1$} & \textcolor{Black}{$-477\pm20$} & \textcolor{Black}{$5492\pm122$} &                              &                              &                              &                             & \textcolor{Black}{$131$} & \textcolor{Black}{$109$} \\

   \hline\hline
   opa2018a & d$X$ & $13\pm1$                      & $381\pm12$                     &                                 & $36\pm1$                     & $-18\pm1$                    &                              &                             & $126$                    & $116$                    \\
            & d$Y$ & $-71\pm1$                     & $-34\pm12$                     &                                 & $-35\pm1$                    & $40\pm1$                     &                              &                             & $147$                    & $128$                    \\
   \hline
   gsf2016a & d$X$ & \textcolor{Black}{$49\pm1$}   & \textcolor{Black}{$441\pm10$}  &                                 & \textcolor{Black}{$43\pm1$}  & \textcolor{Black}{$-11\pm1$} &                              &                             & \textcolor{Black}{$134$} & \textcolor{Black}{$101$} \\
            & d$Y$ & \textcolor{Black}{$-87\pm1$}  & \textcolor{Black}{$61\pm10$}   &                                 & \textcolor{Black}{$-34\pm1$} & \textcolor{Black}{$47\pm1$}  &                              &                             & \textcolor{Black}{$131$} & \textcolor{Black}{$107$} \\
   \hline\hline
   opa2018a & d$X$ & $13\pm1$                      & $334\pm12$                     &                                 & $34\pm1$                     & $-24\pm1$                    & $-19\pm1$                    & $-3\pm1$                    & $126$                    & $116$                    \\
            & d$Y$ & $-70\pm1$                     & $4\pm12$                       &                                 & $-38\pm1$                    & $46\pm1$                     & $22\pm1$                     & $23\pm1$                    & $147$                    & $126$                    \\
   \hline
   gsf2016a & d$X$ & \textcolor{Black}{$49\pm1$}   & \textcolor{Black}{$413\pm11$}  &                                 & \textcolor{Black}{$41\pm1$}  & \textcolor{Black}{$-14\pm1$} & \textcolor{Black}{$-13\pm1$} & \textcolor{Black}{$3\pm1$}  & \textcolor{Black}{$134$} & \textcolor{Black}{$101$} \\
            & d$Y$ & \textcolor{Black}{$-84\pm1$}  & \textcolor{Black}{$40\pm11$}   &                                 & \textcolor{Black}{$-43\pm1$} & \textcolor{Black}{$48\pm1$}  & \textcolor{Black}{$20\pm1$}  & \textcolor{Black}{$23\pm1$} & \textcolor{Black}{$131$} & \textcolor{Black}{$105$} \\
   \hline
  \end{tabular}}
\end{table*}

We fit all of the three models to the residuals and show the results in Table~\ref{tab_VLBI}; the two data \textcolor{Black}{ sources are presented for comparison.} \textcolor{Black}{The uncertainties of the fit parameters are smaller than those presented in \citet{Capitaine2009} because the data have accumulated over ten more years and also because they have a higher quality.}
A parabola clearly is not an effective model to fit the curvature  because the secular and quadratic term are strongly correlated (\textcolor{Black}{$-0.9$}). For the second model, the weighted root mean square (WRMS) is reduced by \textcolor{Black}{about 15\%}. The correlation coefficient between the secular term and the sine term of the 18.6-year nutation \textcolor{Black}{($0.4$) of model 3} stands out among others. \textcolor{Black}{The short time-span of the VLBI observations (38 years) leads to an evident correlation between the 18.6-year nutation term and long-period terms ($t^1$ and $t^2$). All the correlation coefficients can be found in Appendix A.}

By comparing \textcolor{Black}{the} results of models 2 and 3, we can see that there is no significant difference between the fitted coefficients of the 18.6-year term \textcolor{Black}{with and without} the 9.3-year term\textcolor{Black}{. The} correlation coefficients between the two nutation terms are quite small (\char`\~0.2), indicating that \textcolor{Black}{they} are well separated. \textcolor{Black}{The results of CPO series obtained from the two data centers are highly consistent} in terms of nutation terms and uncertainties of all fit coefficients, but \textcolor{Black}{differences can be seen especially in the secular terms of d$Y$. This is probably related to the strong correlation between constant and secular terms ($-0.8$ \textcolor{Black}{in both solutions}) and to the significant uncertainties compared to those of other estimated terms.}

The fitted coefficients of the secular terms show an underestimation \textcolor{Black}{in the IAU model} of the precession rate in \textcolor{Black}{$X$} about $0.3$ \textcolor{Black}{mas cy$^{-1}$}. ~\citet{Capitaine2009} have pointed out that the 18.6-year nutation term is the most sensitive to the error in the precession-nutation model, \textcolor{Black}{with} VLBI data up to 2008. With the accumulated high-precision data, we here obtain more \textcolor{Black}{consistent} results of the corrections of the 18.6-year nutation term. All fit coefficients of the 18.6-year nutation term reveal that the amplitude is underestimated by about $35$ \si{\mu as}. \textcolor{Black}{However, the formal errors of the CPO derived from VLBI data are probably underestimated by about a factor 2 according to ~\citet{Herring2002}. }
\section{LLR observations and CPO series }
\label{section_LLR}
The LLR observations are presented as so-called normal points. They refer to lines of data that contain the emission time of the laser, the observed round trip time in UTC, the telescope and reflector ID, and some atmospheric parameters of each observation. These data can be used to calculate the round-trip times and then the residuals of the round-trip time [observation \textcolor{Black}{minus} calculation (O$-$C) ], which can be converted into residuals in one-way distance in centimeters. Finally, we obtain CPO series based on these residuals.
\subsection{LLR data}
We used the residuals \textcolor{Black}{spanning} 1970-\textcolor{Black}{2017} (O$-$C of the one-way distance in centimeters) provided by~\citet{Pavlov2016}\footnote{\url{http://iaaras.ru/en/dept/ephemeris/llr-oc/}}. The rejection procedure is taken in advance to exclude data with relatively poor quality. The rejection criterion is such that \textcolor{Black}{ O$-$C values that are higher than three times  the respective formal error and the total WRMS of all normal points of the same station are excluded}. The numbers of the used and rejected normal points and the corresponding time coverages of each LLR ground-based station are listed in Table~\ref{LLR data count}. Data of \textcolor{Black}{the McDonald Laser Ranging Station (MLRS2)} between 2000 and \textcolor{Black}{2015} have a relatively lower accuracy and accordingly cause lower accuracies in the CPO \textcolor{Black}{that were obtained between 2000 and 2005 (the WRMS of MLRS2 in this time span is 12.18 cm, while the overall WRMS of this time span is 4.54 cm)}  in the following sections.
\begin{table*}[!htbp]
 \caption{Basic information of LLR data of each station.}
 \label{LLR data count}
 \centering
 \begin{tabular}{lrrrr}
  \hline\hline
  station name & observation duration                              & WRMS(cm) & N$_{\rm{total}}$ & N$_{\rm{rejected}}$ \\
  \hline
  APOLLO       & \textcolor{Black}{2006}.04.07 — 2016.11.25 & 1.03     & 2648             & 336                 \\
  Haleakala    & 1984.11.13 — 1990.08.30                           & 5.40     & 770              & 202                 \\
  Matera       & 2003.02.22 —2017.11.10                            & 8.44     & 105              & 26                  \\
  McDonald     & 1970.07.20 — 1985.06.30                           & 77.54    & 3575             & 117                 \\
  MLRS1        & 1983.08.02 — 1988.01.27                           & 40.64    & 631              & 58                  \\
  MLRS2        & 1988.02.29 — 2015.03.25                           & 7.41     & 3669             & 510                 \\
  OCA(IR)      & 2015.03.11 — 2017.12.21                           & 0.92     & 2839             & 105                 \\
  OCA(MeO)     & 2009.11.11 — 2017.12.21                           & 1.33     & 1836             & 32                  \\
  OCA(Ruby)    & 1984.06.11 — 1986.06.12                           & 36.58    & 1112             & 3                   \\
  OCA(YAG)     & 1987.10.12 — 2005.07.30                           & 7.27     & 8316             & 493                 \\
  \hline
  Total        & 1970.07.20 — 2017.12.21                           & 2.06     & 25501            & 1882                \\
  \hline
 \end{tabular}
\end{table*}

\subsection{General methods of calculating CPO from LLR \textcolor{Black}{residual}}
\label{methodLLR}
ZC09 offered a method of calculating the \textcolor{Black}{CPO} from LLR \textcolor{Black}{residual}. We generally followed their method \textcolor{Black}{in converting LLR residuals into CPO} \textcolor{Black}{(i.e., Eqs.~(\ref{D}) to~(\ref{partial})). \textcolor{Black}{The methods of deriving LLR residuals from the observational data (also known as ``normal points'') in ZC09~\citep{Chapront1999} and ~\citet{Pavlov2016} are basically the same. It is noted in both methods that} not separating the two ways of light travel in the calculation of the round-trip time may cause errors of hundreds of meters in residuals of the one-way distance. Therefore, \textcolor{Black}{different from   ZC09,} we} \textcolor{Black}{also} treated the forward and backward ways of light travel separately \textcolor{Black}{in this part} for a more solid estimation \textcolor{Black}{(further discussion in Sect.~\ref{two-way}), namely taking} the round-trip time as \textcolor{Black}{$\Delta t=(\vec D_1+\vec D_2)/c$} instead of $\Delta t=2\vec D/c$, in which $\vec D_1$ and $\vec D_2$ are the station-reflector vectors of the forward and backward ways of light travel in the barycentric celestial reference system (BCRS).

The station-reflector vector can be written as the summation of three vectors, as presented in ZC09,
\begin{equation}\label{D}
 \vec{D}=\vec{SE}+\vec{EM}+\vec{MR},
\end{equation}
where $-\vec{SE}$ is the geocentric position of the ground-based station, $\vec{EM}$ is the geocentric position of the moon, \textcolor{Black}{and} $\vec{MR}$ is the selenocentric position of the reflector. In order to calculate the \textcolor{Black}{partial derivative of $\Delta t$ with respect to the} celestial pole coordinate $X$ and $Y$, we need to use the transformation from ITRS to GCRS for the station coordinates:
\begin{equation}
 \vec {SE}_{\rm{GCRS}}=\mathcal Q(t)\mathcal R(t)\mathcal W(t)\bm{SE}_{\rm{ITRS}}=\mathcal{M}(t)\bm{SE}_{\rm{ITRS}}.
\end{equation}
In the expression above, only $\mathcal Q$, the transformation matrix for precession and nutation, is related to $X$ and $Y$:
\begin{equation}
 \mathcal Q=\mathcal{PN}(X,Y)\mathcal R_{\rm{z}}(s),
\end{equation}
where $\mathcal{PN}$ is the precession nutation matrix and $s$ is the CIO locator. Thus the partial derivatives of the calculated time delay with respect to $X$ and $Y$ \textcolor{Black}{can be} written as
\begin{equation}
 \label{partial}
 \begin{split}
  &\frac{\partial\Delta t}{\partial X}=\frac{1}{c}\left[\frac{\vec D_1}{D_1}\left(-\frac{\partial\mathcal M}{\partial X}(t_1,t_2)\vec{ES}_{\rm{ITRS}}\right)+\frac{\vec D_2}{D_2}\left(-\frac{\partial\mathcal M}{\partial X}(t_3,t_2)\vec{ES}_{\rm{ITRS}}\right)\right] \\
  &\frac{\partial\Delta t}{\partial Y}=\frac{1}{c}\left[\frac{\vec D_1}{D_1}\left(-\frac{\partial\mathcal M}{\partial Y}(t_1,t_2)\vec{ES}_{\rm{ITRS}}\right)+\frac{\vec D_2}{D_2}\left(-\frac{\partial\mathcal M}{\partial Y}(t_3,t_2)\vec{ES}_{\rm{ITRS}}\right)\right],
 \end{split}
\end{equation}
in which $t_1$ is the emission time, $t_2$ is the reflection time, and $t_3$ is the reception time. We follow Sect. 4 in~\citet{Pavlov2016} to calculate $t_2$ and $t_3$, but ignore some minor effects such as solid tide deformations of the Earth and the Moon because we do not need an accuracy of \textcolor{Black}{ the moment of time} as high as that in the calculation residuals of the one-way distance. The relativistic transformations between terrestrial and selenocentric reference systems to the BCRS~\citep{Petit2010} were applied to transform the vectors in Eq.~(\ref{D}) into BCRS. The relativistic gravitational delay~\citep{Kopeikin1990} and the tropospheric delay~\citep{Mendes2004,Mendes2002} were also taken into account.
We used Jet Propulsion Laboratory (JPL) planetary ephemerides DE430~\citep{Folkner2014} for the geocentric and barycentric positions of the Sun, Earth, Moon, and major planets, and also for the lunar \textcolor{Black}{libration} angles and TT$-$TDB transformations.

After obtaining the partial derivatives in Eq.~(\ref{partial}), we derived $\mathrm{d\Delta}t$ by dividing twice the residuals in one-way distance with the speed of light in vacuum. Then we obtained the CPO by weighted least-square fits using
\begin{equation}
 \begin{split}
  & \mathrm{d\Delta}t=\frac{\partial\Delta t}{\partial X}\mathrm{d}X+\frac{\partial\Delta t}{\partial Y}\mathrm{d}Y.
 \end{split}
\end{equation}

\subsection{Uncertainties of \textcolor{Black}{the} determined CPO}
Uncertainties are presented as two and three times the formal error in ZC09 and H18, respectively, after fitting the CPO series to the models to account for unmodeled effects and further model deficiencies. \textcolor{Black}{The factor three was checked by analyses of subsets of \textcolor{Black}{used} LLR normal points.} Here we present an estimation of uncertainties in advance of obtaining CPO series from LLR normal points.

\textcolor{Black}{We investigated the uncertainty of the CPO derived from LLR observations, considering that}  many aspects would add to the uncertainty, except for the observational formal errors. According to IERS conventions~\citep{Petit2010}, the accuracy of satellite and lunar laser ranging (SLR and LLR) is greatly affected by the residual errors in modeling the effect of signal propagation through the troposphere and stratosphere. The traditional approach in laser ranging data \textcolor{Black}{analyses} used a model developed in the 1970s~\citep{Marini1973}, until new mapping functions~\citep{Mendes2002} and the zenith delay model~\citep{Mendes2004} were used after January 1, 2007.~\citet{Mendes2003} assessed the ``model minus ray tracing (cm)'' for several wavelengths used in LLR and SLR techniques with the elevation angle set to be $\ang{10}$ for both models mentioned above. For the typical wavelength used in the LLR technique ($532$ nm), the assessed ``model minus ray tracing (cm)'' was $0.82$ cm for the currently used models.

Moreover, uncertainties of orbit and \textcolor{Black}{libration} determinations of the Moon will also add to that of the calculated CPO. We estimated the influence of these aspects by using different ephemerides in our calculation of the CPO and comparing the results. We \textcolor{Black}{applied} two different ephemerides (DE430 and INPOP17a \textcolor{Black}{~\citep{Viswanathan2017} from IMCCE of Paris Observatory}) to the calculation of the CPO series to estimate the level of differences in residuals in the one-way distance. The weighted mean of the resulting differences in the one-way distance \textcolor{Black}{is} $0.11$ cm (DE430$-$INPOP17a).

Uncertainty estimates of the calculation process can only be obtained in a statistical way as presented above because the process of determining residuals of LLR observations \textcolor{Black}{is} extremely complicated and \textcolor{Black}{prevents} us from obtaining actual uncertainties for each normal point. The sum of the two estimates above is approximately twice the formal error of LLR observations ($0.61$ cm).

Because of the complexity and according to the results of statistical tests above, we multiplied the \textcolor{Black}{formal errors of residual} in the one-way distance with a factor three before we obtained the CPO with weighted least-square fits \textcolor{Black}{to compensate for the mixture of systematic and random errors}. In this way, the \textcolor{Black}{weighted} average uncertainty of CPO determined from LLR observations is \textcolor{Black}{$0.061$ mas and is almost comparable with VLBI}\textcolor{Black}{, but we stress that the quality of the CPO series also depends on the resolution (\char`\~70 days for LLR instead of 1 day for VLBI), which appears to be another shortcoming of LLR that is due to the poor time distribution of the observation.}

\begin{figure}[!htbp]
 \centering
 \includegraphics[width=8cm]{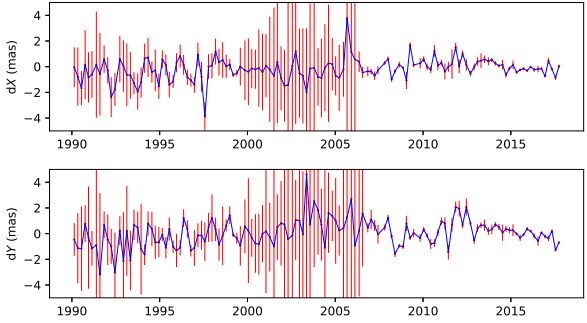}
 \caption{Details after 1985 of d$X$ and d$Y$ residuals from LLR observations, using separated windows of 70 days.}
 \label{CPO_LLR_70d_85}
\end{figure}

The fitting interval of each CPO was chosen as 70 days as in ZC09 for the whole time span of LLR observations. Finally, we obtained a series of 220 corrections (d$X$ and d$Y$) out of 23619 normal points. The obtained CPO series is plotted in Fig.~\ref{CPO_LLR_70d}, and details after 1985 are presented in Fig.~\ref{CPO_LLR_70d_85}.

\subsection{Analyses of LLR residuals}
\label{analysis_LLR}
\begin{table*}[!htbp]
 \caption{Weighted fits of models 2 and 3 to LLR residuals.}
 \label{tab_LLR}
 \centering
 \scalebox{0.9}{
  \begin{tabular}{r r r r r r r r r r}
   \hline\hline
             \textcolor{Black}{Term}&                     & $t^0$          & $t^1$          & sin(18.6 yr)   & cos(18.6 yr)   & sin(9.3 yr)    & cos(9.3 yr)    & WRMS$_{\rm{pre}}$ & WRMS$_{\rm{post}}$ \\
             \textcolor{Black}{Unit}&                     & mas            & mas/cy         & mas            & mas            & mas            & mas            & mas               & mas                \\
   \hline
   This work & d$X$                & $-0.38\pm0.02$ & $1.43\pm0.18$  & $-0.26\pm0.02$ & $-0.37\pm0.01$ &                &                & $0.526$           & $0.463$            \\
             & d$Y$                & $-0.36\pm0.03$ & $-0.54\pm0.19$ & $-0.81\pm0.02$ & $-0.30\pm0.01$ &                &                & $0.672$           & $0.581$            \\
   \hline
   ZC09      & d$X$                & $0.27\pm0.13$  & $5.77\pm3.25$  & $0.00\pm0.22$  & $0.01\pm0.13$  &                &                                                         \\
             & d$Y$                & $-0.17\pm0.13$ & $1.07\pm3.11$  & $-0.02\pm0.21$ & $-0.22\pm0.12$ &                &                                                         \\
   \hline\hline
   This work & d$X$                & $-0.28\pm0.03$ & $1.77\pm0.19$  & $-0.14\pm0.04$ & $-0.26\pm0.02$ & $0.20\pm0.02$  & $0.01\pm0.02$  & $0.526$           & $0.458$            \\
             & d$Y$                & $-0.12\pm0.04$ & $0.02\pm0.19$  & $-0.25\pm0.05$ & $-0.24\pm0.02$ & $-0.05\pm0.02$ & $-0.40\pm0.03$ & $0.672$           & $0.562$            \\
   \hline
   ZC09      & d$X$                & $0.16\pm0.15$  & $3.52\pm3.84$  & $0.17\pm0.27$  & $0.12\pm0.14$  & $0.12\pm0.16$  & $0.32\pm0.14$  &                   &                    \\
             & d$Y$                & $-0.22\pm0.14$ & $-0.16\pm3.67$ & $0.08\pm0.26$  & $-0.24\pm0.14$ & $0.10\pm0.15$  & $-0.01\pm0.14$ &                   &                    \\
   \hline
   H18       & d$\psi\sin\epsilon$ &                &                & $0.58\pm0.18$  & $-0.09\pm0.13$ & $0.04\pm0.12$  & $-0.01\pm0.12$ &                   &                    \\
             & d$\epsilon$         &                &                & $-0.12\pm0.17$ & $-0.36\pm0.16$ & $-0.49\pm0.12$ & $0.17\pm0.13$  &                   &                    \\
   \hline
  \end{tabular}}
\end{table*}
We present estimates using the empirical models 2 and 3 for the long-period components of nutation after the FCN was removed (\textcolor{Black}{cf. Sect.~\ref{section_VLBI} for VLBI data}). Results of the previous works are listed for comparison in Table~\ref{tab_LLR}. \textcolor{Black}{ H18 presented the results as in-phase and out-of-phase terms of $\mathrm{d}\psi$ and $\mathrm{d}\epsilon$, and we converted them into $\mathrm{d}\psi\sin\epsilon$ and $\mathrm{d}\epsilon$.} The decrease in WRMS after the weighted least-square fits is clear \textcolor{Black}{(about 20\%) }, and the largest deviations of all nutation terms occur for the 18.6-year sine term, \textcolor{Black}{in accordance} with the results of ZC09 and H18. The fit coefficients show no evident correction to the precession rate, but a deviation of about $-0.3$ mas for all of the coefficients of the 18.6-year nutation term. The determined accuracies are better than those in ZC09 because of ten more years of observational time coverage. Furthermore,~\citet{Pavlov2016} treated the LLR data meticulously, including a discussion of the limits of the widely used IERS C04 series (see details in Sect. 4 in their paper), \textcolor{Black}{providing us with a O$-$C series of high quality,} which also contributes to the better accuracy.

\textcolor{Black}{H18 presented an overall estimation with LLR data from 1970 to 2016. Five nutation terms were estimated along with the station coordinates, velocities, and retro-reflector coordinates, and thus uncertainties were affected by the correlations. For instance, correlations were up to 0.5 between nutation coefficients and the perturbation rotation rate in y-direction, up to 0.4 between the sine term of $\mathrm{d}\psi$ and cosine term of $\mathrm{d}\epsilon$ with the $z$-component of the retro-reflector coordinates, and up to 0.7 between the cosine term of $\mathrm{d}\psi$ and sine term of $\mathrm{d}\epsilon$ with the initial $z$-components of the lunar position. In our method, however, \textcolor{Black}{these values are} fixed by applying the ephemeris DE430 and the coordinates of the stations and retro-reflectors estimated by ~\citet{Pavlov}.}

In our results, the fit coefficients of the 18.6-year nutation term \textcolor{Black}{with and without the 9.3-year term} are not consistent. This feature is different from the results obtained by VLBI analyses. \textcolor{Black}{Meanwhile, the correlation coefficients between two nutation terms are over 0.7, revealing the incapability of LLR data to separate the components effectively. This is probably because that LLR observations are directly related with the motions of the Moon, which is also the most important excitation of the 18.6-year and 9.3-year nutations. Nevertheless, the correlation coefficients between the secular term and the nutation terms are generally smaller than those of VLBI, probably benefiting from the longer time span of ten years. Furthermore, the correlation coefficient between the secular term and the sine term of 18.6yr nutation remains larger than its counterparts, which may reveal a common problem shared by the VLBI and LLR techniques.}
\subsection{Influence of not separating the two \textcolor{Black}{ways} of light travel in LLR calculation}
\label{two-way}
\textcolor{Black}{As explained in Sect.~\ref{methodLLR}, we separated the two ways of light travel when we converted LLR residuals into CPO. We tested the effect of this factor} by using only the emission time when we calculated the partial derivatives (this equals the first iteration in the calculation process used in this work), and placed the external residuals~\citep{Pavlov2016} into the least-squares fit to obtain d$X$ and d$Y$. Then we used this CPO series to estimate the same corrections for the precession and nutation terms. The results only differed by $10^{-5}$ of their formal errors compared to the results in Sect.~\ref{analysis_LLR}. This indicates that the Earth rotation parameters are not as sensitive to the accuracies of reflection and reception times as the laser ranging itself.
\begin{figure}[!htbp]
 \centering
 \includegraphics[width=8cm]{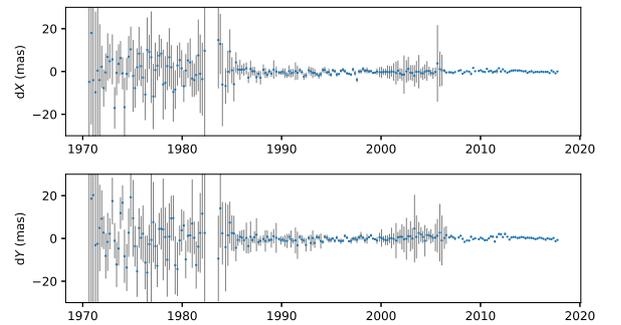}
 \caption{d$X$ and d$Y$ residuals of LLR observations in mas, using separated windows of 70 days.}
 \label{CPO_LLR_70d}
\end{figure}
\section{CPO analyses with optical observations}
\label{section_optical}
The history of optical observations can be traced back to the nineteenth century: this can be considered as an advantage in determining long-term effects in the Earth rotation. We used the OA00 series constructed by~\citet{Vondrak2010}, \textcolor{Black}{which is} based on a combination of the Hipparcos and Tycho catalogs and long-lasting ground-based astrometric observations. It was derived from the Earth orientation catalogs (EOC) containing 4418 celestial objects. The catalogs were used to derive the CPO during the time interval 1899.7-1992.0, referred to as the Hipparcos celestial reference frame (HCRF).
\subsection{Transformation of optical data}
The OA00 series \textcolor{Black}{provides} time series of nutation offsets d$\psi$ and d$\epsilon$ with respect to the IAU\,1976/1980 precession-nutation models~\citep{Lieske1977,Wahr1981,Seidelmann1982} \textcolor{Black}{between 1899 and 1972}. Therefore, it is necessary to transform them into the same system as the VLBI series, that is, d$X$ and d$Y$ with respect to the IAU\,2006/2000 precession-nutation models.

\begin{figure}[!htbp]
 \centering
 \includegraphics[width=8cm]{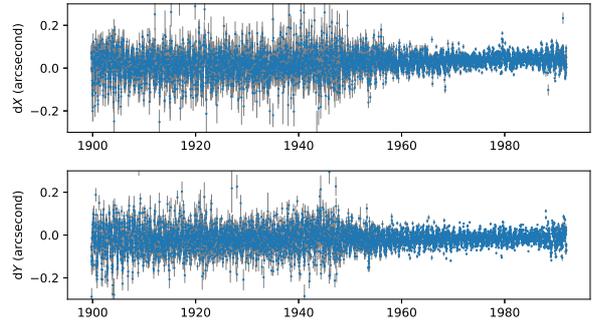}
 \caption{d$X$ and d$Y$ residuals from optical observations.}
 \label{fig_OA00}
\end{figure}

\textcolor{Black}{We followed the method described by~\citet{Capitaine2006}.} First we evaluated the observed nutation angles of date by adding the optical nutation offsets to the IAU\,1980 nutation model,
\begin{equation}
 \Delta\psi=\Delta\psi_{\rm{IAU1980}}+\mathrm{d}\psi, \Delta\epsilon=\Delta\epsilon_{\rm{IAU1980}}+\mathrm{d}\epsilon.
\end{equation}
Then we calculated the nutation matrix with $\Delta\psi$, $\Delta\epsilon,$ and the mean obliquity of date ($\epsilon_A$) by
\begin{equation}
 \mathcal{N}=  \mathcal{R_{\rm 1}}(-\epsilon_A)\mathcal{R_{\rm 3}}(\Delta\psi)\mathcal{R_{\rm 1}}(\epsilon_A+\textcolor{Black}{\Delta}\epsilon).
\end{equation}
\textcolor{Black}{We applied the precession matrix (IAU\,1976) of date to form the precession-nutation matrix $\mathcal M_{\rm obs}$}, of which the elements (3,1) and (3,2) are the observed $X$ and $Y$ coordinates of the CIP in the GCRS. Finally, we obtained the celestial pole offsets with respect to the IAU\,2006/2000 model:
\begin{equation}
 \begin{split}
  \mathrm{d}X=\mathcal M_{\rm obs}(3,1)-X_{\rm{IAU2006/2000}},\\
  \mathrm{d}Y=\mathcal M_{\rm obs}(3,2)-Y_{\rm{IAU2006/2000}}.
 \end{split}
\end{equation}
The obtained CPO in arcseconds is plotted in Fig.~\ref{fig_OA00}.
\subsection{Analyses of the optical residuals}
Because the OA00 series has \textcolor{Black}{a} resolution of 5 days~\citep{Vondrak2000}, it is theoretically able to determine the long-periodic \textcolor{Black}{nutation terms}.~\citet{Vondrak2000} have performed certain analyses of the residual series OA97 and OA99, the predecessors of OA00 and OA10, at the end of the 1990s, and compared the results to VLBI observations back then. According to their results, the fit results of OA99 were closer to those of the VLBI series than OA97. The accuracies of fit coefficients of OA99 and VLBI were on the same order owing to the low quality of VLBI data before 1990. Since then, VLBI observations have lasted 20 more years and have been greatly improved. The corresponding accuracy of the fit coefficients of corrections to long-period terms in precession-nutation models has reached \si{\mu as} level, leaving the optical data an only advantage of the long history. \textcolor{Black}{The uncertainties of the CPO derived from OA00 series are $11.33$ and $7.66$ mas (weighted average) in d$X$ and d$Y,$ respectively, about two hundred times that of VLBI data.}

\textcolor{Black}{Models 1 and 2 were fit to the transformed CPO} to estimate the possible long-term corrections. The results are presented in Table~\ref{tab_OA00}. In both models $t$ refers to the number of centuries from a reference epoch J1956.0, which is almost the central epoch of the time coverage of optical observations. In contrast to the VLBI results, there is no significant difference between the constant and secular terms in these two tables, \textcolor{Black}{ and the correlation coefficients between secular term and nutation terms are negligible. This indicates that a time span long enough (over ninety years) may be sufficient to separate the quadratic term and the 18.6-year nutation term, whereas optical data cannot help in improving the precession-nutation model today because of the poor accuracy.}

\begin{table*}[!htbp]
 \caption{Weighted fits of models 1 and 2 to optical residuals (1899.7-\textcolor{Black}{1991.9}) corresponding to the IAU\,2006/2000 model.}
 \label{tab_OA00}
 \centering
 \begin{tabular}{l r r r r r r r}
  \hline\hline
                             \textcolor{Black}{Term}& $t^0$                                   & $t^1$         & $t^2$           & sin(18.6 yr) & cos(18.6 yr)  & WRMS$_{\rm{pre}}$                & WRMS$_{\rm{post}}$ \\
                             \textcolor{Black}{Unit}& mas                                     & mas/cy        & mas/$\rm{cy}^2$ & mas          & mas           & mas                              & mas                \\
  \hline
  $\Delta\psi\sin\epsilon_A$ & $19.8\pm0.3$                            & $-72.8\pm0.8$ & $43.1\pm3.3$    &              &               & $35.6$                           & $29.3$             \\
  $\Delta\epsilon$           & $-7.1\pm0.3$                            & $-7.8\pm0.8$  & $34.7\pm3.3$    &              &               & $29.8$                           & $29.1$             \\
  \hline
  d$X$                       & \textcolor{Black}{$-14.8\pm0.3$} & $43.6\pm0.8$  & $49.8\pm3.3$    &              &               & \textcolor{Black}{$31.3$} & $29.5$             \\
  d$Y$                       & \textcolor{Black}{$-12.7\pm0.2$} & $16.0\pm0.6$  & $23.8\pm2.3$    &              &               & \textcolor{Black}{$30.3$} & $28.7$             \\
  \hline\hline
  d$X$                       & \textcolor{Black}{$-11.3\pm0.2$} & $40.2\pm0.8$  &                 & $1.95\pm0.2$ & $0.06\pm0.2$  & \textcolor{Black}{$31.3$} & $29.6$             \\
  d$Y$                       & \textcolor{Black}{$-11.2\pm0.1$} & $13.97\pm0.6$ &                 & $0.82\pm0.2$ & $-1.58\pm0.2$ & \textcolor{Black}{$30.3$} & $28.7$             \\
  \hline
 \end{tabular}
\end{table*}
\section{Joint analyses of VLBI and LLR residuals}
\textcolor{Black}{Because the observing history of VLBI is short and the optical and LLR data lack high accuracy, a combined series would benefit from the advantages of each technique while avoiding shortcomings. However, the quality of the optical data is far too poor compared with the other two techniques and therefore can barely contribute anything. Therefore,} we only performed \textcolor{Black}{joint analyses} of VLBI (OPA series) and LLR observations.
\label{section_joint}
\subsection{Systematic deviations between LLR and VLBI systems}
It should be noted that the reference frame for LLR observations \textcolor{Black}{is} dynamically defined and the reference frame for VLBI \textcolor{Black}{is}  kinematic, so that it \textcolor{Black}{is} necessary to take into account the small systematic deviations between these two reference systems before the two series are combined.~\citet{Liu2012} has developed the relationships between relative orientation offset with a rigid-body rotation $(\epsilon_{\rm x}, \epsilon_{\rm y}, \epsilon_{\rm z})$ for two reference frames and the difference of EOPs $(\Delta X, \Delta Y)$ expressed in these two frames:
\begin{equation}
 \begin{split}
  &\mathrm{\Delta}X=+\epsilon_{\rm z}Y-\epsilon_{\rm y}Z\\
  &\mathrm{\Delta}Y=-\epsilon_{\rm z}X+\epsilon_{\rm x}Z ,
 \end{split}
\end{equation}
in which $\epsilon_{\rm x}$, $\epsilon_{\rm y}$, and $\epsilon_{\rm z}$ are the rotation angles about the three axes. In the case of short time-intervals (several hundred years), the CIP coordinates $X$ and $Y$ are small quantities, and thus $Z=\sqrt{1-X^2-Y^2}\approx1$, so that the effect on EOP can be limited to the first order in $X$, $Y$, and $\epsilon_{\rm x,y,z}$:
\begin{equation}
 \begin{split}
  &\mathrm{\Delta}X\approx-\epsilon_{\rm y} \\
  &\mathrm{\Delta}Y\approx+\epsilon_{\rm x}.
 \end{split}
\end{equation}
With this method, we obtained the upper limit of the bias between the dynamical and kinematic reference frames by calculating weighted means of the differences between CPO series obtained from VLBI and LLR observations. The uncertainties were derived as the average of
\begin{equation}
 \sigma=\sqrt{\sigma_{\rm LLR}^2+\sigma_{\rm VLBI}^2},
\end{equation}
representing the accuracy of LLR and VLBI observations. The results are
\begin{equation}
 \begin{split}
  &\epsilon_{\rm x}=16.42\pm210.50\,\si{\mu as}\\
  &\epsilon_{\rm y}=-148.37\pm195.02\,\si{\mu as}.
 \end{split}
\end{equation}
Because the derived uncertainties are \textcolor{Black}{larger} than the biases themselves, we consider the two systems to be consistent.
\begin{figure*}[h]
 \centering
 \includegraphics[width=0.8\linewidth]{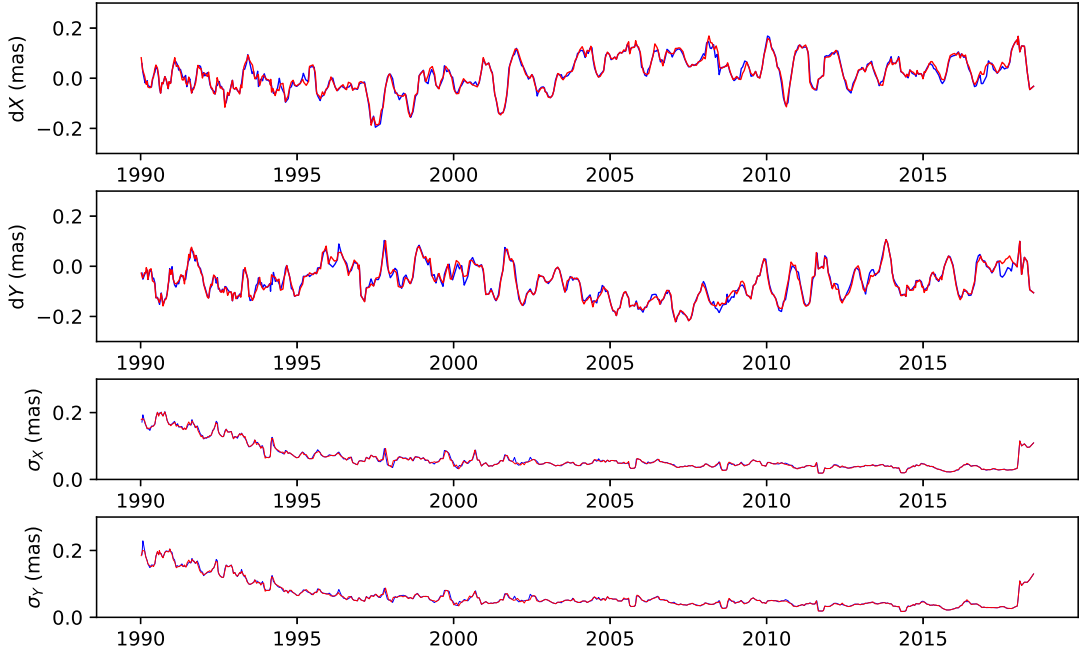}
 \caption{Combined series of VLBI and LLR observations (blue) and the mean VLBI series obtained with 70-day sliding windows (red).}
 \label{joint-fig}
\end{figure*}
\subsection{Details of the combined series}
The method of combination is \textcolor{Black}{almost the same as that applied in ~\citet{Zerhouni2009} (Sect. 7): obtaining weighted means} within sliding windows of 70 days and sliding every 10 data points, after removing the FCN from the opa2018a and LLR series. We obtained a combined series containing 655 points with this method. We also obtained a weighted mean series of VLBI (also \textcolor{Black}{with} 70-day sliding windows) for comparison, and it contains 633 points. The combined series is plotted together with the series of mean CPO from VLBI in each window (see Fig.~\ref{joint-fig}).

\textcolor{Black}{The features of the combined series are almost the same as that of the smoothed VLBI series, considering the better quality of the VLBI data. The differences in the figure indicate that the LLR CPO series has a comparable accuracy in certain periods (e.g., 1993-1995 and 2007-2010) that possibly reveals different details.} The fitting results to all three models are listed in Table.~\ref{tab_joint}. The coefficients are slightly different from those in Table.~\ref{tab_VLBI}. The WRMS of the combined series \textcolor{Black}{reduces} due to data smoothing in the combining method.
\begin{table*}[!htbp]
 \caption{Weighted fits of the joint series (1970.7-2018.5) to three models, corresponding to the IAU\,2006/2000 model.}
 \label{tab_joint}
 \centering
 \scalebox{0.9}{
  \begin{tabular}{rrrrrrrrrrr}
   \hline\hline
        \textcolor{Black}{Term}& $t^0$     & $t^1$       & $t^2$         & sin(18.6 yr) & cos(18.6 yr) & sin(9.3 yr) & cos(9.3 yr) & WRMS$_{\rm{pre}}$ & WRMS$_{\rm{post}}$ \\
        \textcolor{Black}{Unit}& mas       & mas/cy      & mas/cy$^2$      & mas          & mas          & mas         & mas         & mas               & mas                \\
   \hline
   d$X$ & $18\pm3$  & $286\pm56$  & $-6606\pm380$ &              &              &             &             & $71$              & $60$               \\
   d$Y$ & $-83\pm3$ & $-510\pm58$ & $5024\pm388$  &              &              &             &             & $91$              & $60$               \\
   \hline
   d$X$ & $12\pm3$  & $371\pm31$  &               & $36\pm3$     & $-22\pm3$    &             &             & $71$              & $53$               \\
   d$Y$ & $-70\pm3$ & $-61\pm32$  &               & $-37\pm3$    & $38\pm3$     &             &             & $91$              & $55$               \\
   \hline
   d$X$ & $13\pm3$  & $338\pm32$  &               & $35\pm3$     & $-27\pm3$    & $-15\pm3$   & $-4\pm3$    & $71$              & $52$               \\
   d$Y$ & $-70\pm3$ & $-17\pm33$  &               & $-39\pm3$    & $44\pm3$     & $22\pm3$    & $18\pm3$    & $91$              & $52$               \\
   \hline
  \end{tabular}}
\end{table*}
\section{Discussion}
\label{section_discuss}
\subsection{Possibilities of obtaining more CPO corrections out of LLR observations}
In Sect.~\ref{section_LLR} we obtained only 220 corrections \textcolor{Black}{from} 23619 normal points. We adoptedd two methods to explore the possibility of obtaining more corrections from the LLR normal points.
\begin{figure}[!htbp]
 \centering
 \includegraphics[width=8cm]{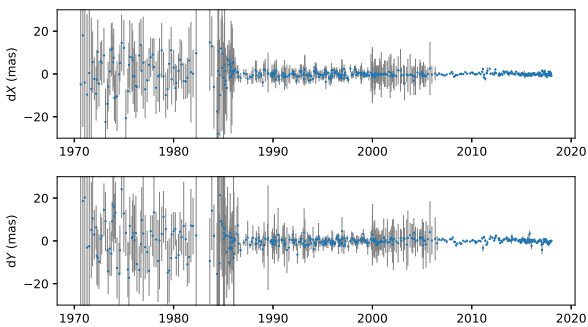}
 \caption{d$X$ and d$Y$ residuals from LLR observations, using changing windows of 50 counts and within 70 days.}
 \label{CPO_LLR_50n}
\end{figure}

\begin{figure}[h]
 \centering
 \includegraphics[width=8cm]{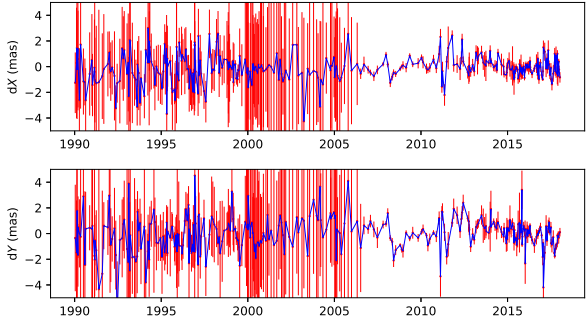}
 \caption{\textcolor{Black}{Details after 1985 of} d$X$ and d$Y$ residuals from LLR observations, using changing windows of 50 counts and within 70 days. \textcolor{Black}{(While it is stated that setting the number of normal points to 50, it appears that there may be 67 windows (out of 483, i.e., more than 10\%) that contain fewer than 50 points, in which case the window length is about 70 days.)}}
 \label{CPO_LLR_50n_85}
\end{figure}

\textcolor{Black}{First,} because the observations are much more frequent after 1985 (see Fig.~\ref{fig-distribution}), it is \textcolor{Black}{possible} to obtain more CPO data points out of the LLR normal points from this period of time by shortening the interval within which one point of the CPO (d$X$ and d$Y$) is derived. We decided to choose the length of the window according to the frequency of observation\textcolor{Black}{. We basically fixed} the number of normal points to be 50 in each window, and the longest \textcolor{Black}{time span of the windows} was set to be 70 days. By doing so, we obtained 67 windows that contained fewer than 50 normal points in periods without observations. The total number of windows is 483, more than twice the series presented in Sect.~\ref{section_LLR} (see in Fig.~\ref{CPO_LLR_50n}. Details \textcolor{Black}{after 1985} are shown in Fig.~\ref{CPO_LLR_50n_85}).

\begin{figure}[h]
 \centering
 \includegraphics[width=8cm]{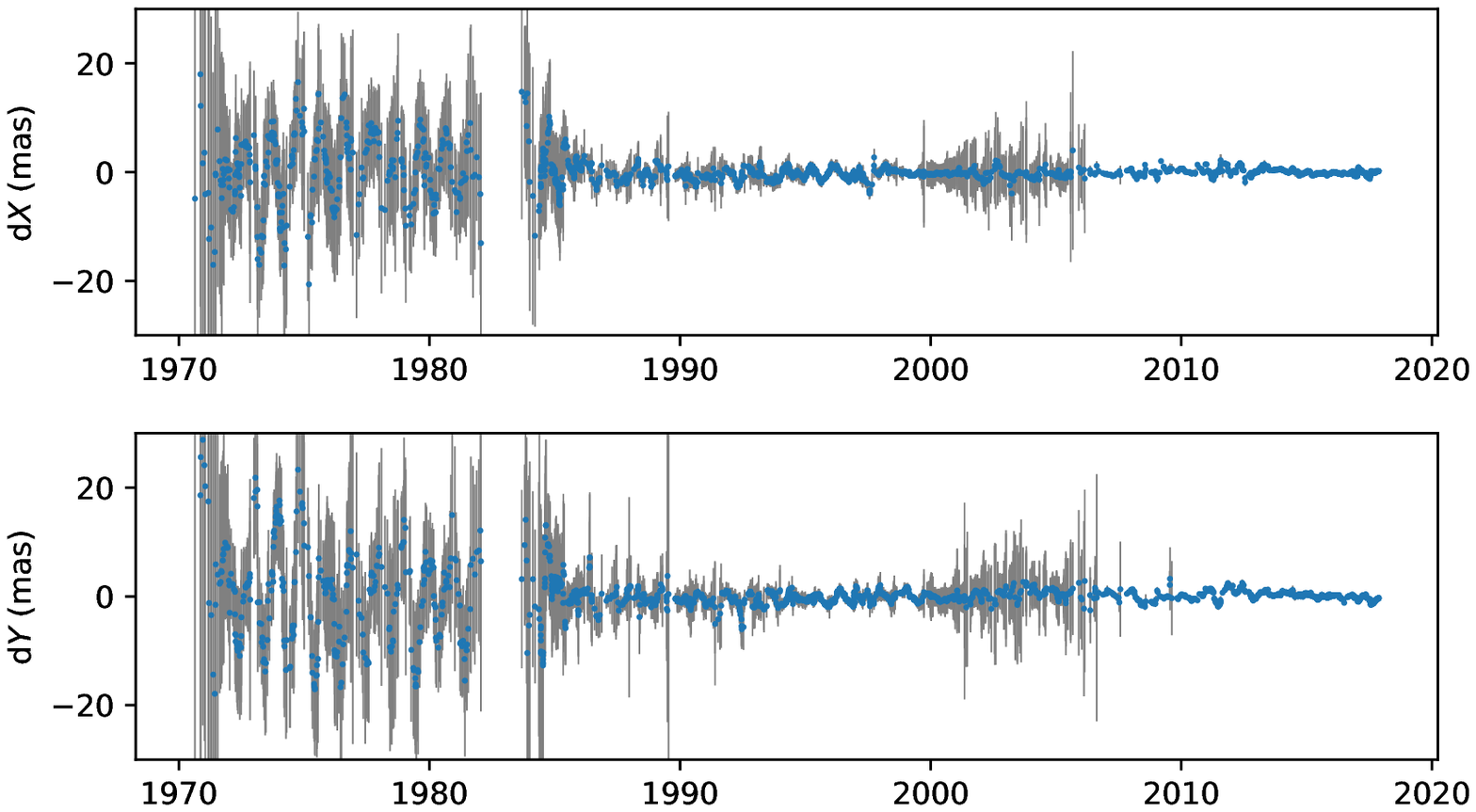}
 \caption{d$X$ and d$Y$ residuals from LLR observations, using sliding windows of 70 days.}
 \label{CPO_LLR_s70}
\end{figure}

\begin{figure}[h]
 \centering
 \includegraphics[width=8cm]{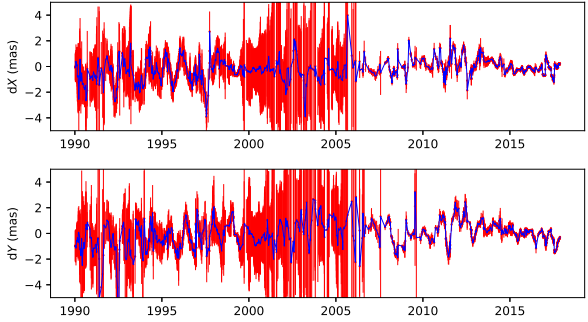}
 \caption{\textcolor{Black}{Details after 1985 of} d$X$ and d$Y$ residuals from LLR observations, using sliding windows of 70 days.}
 \label{CPO_LLR_s70_85}
\end{figure}

Another way to obtain more \textcolor{Black}{CPO data points} is using a sliding window instead of separated windows. We took 70 days as the window length as in Sect.~\ref{section_LLR} and caused the window to slide at eevery 10 normal points. We obtadin 2513 corrections with this method, and \textcolor{Black}{the accuracy appears} to be as good as that of the method using separated windows in Sect.~\ref{section_LLR} (see in Fig.~\ref{CPO_LLR_s70}, and details after 1985 are shown in Fig.~\ref{CPO_LLR_s70_85}).


\begin{table*}[!htbp]
 \caption{Weighted fits of model 3 to LLR residuals obtained with sliding-window method and changing-window method, corresponding to the IAU\,2006/2000 models.}
 \label{LLR-window_test}
 \centering
 \scalebox{0.9}{
  \begin{tabular}{r r r r r r r r r r}
   \hline\hline
            \textcolor{Black}{Term}&      & $t^0$          & $t^1$          & sin(18.6 yr)   & cos(18.6 yr)   & sin(9.3 yr)    & cos(9.3 yr)    & WRMS$_{\rm{pre}}$ & WRMS$_{\rm{post}}$ \\
            \textcolor{Black}{Unit}&      & mas            & mas/cy         & mas            & mas            & mas            & mas            & mas               & mas                \\
   \hline
   sliding  & d$X$ & $-0.19\pm0.01$ & $1.68\pm0.05$  & $-0.02\pm0.01$ & $-0.14\pm0.01$ & $0.27\pm0.01$  & $-0.04\pm0.01$ & $0.460$           & $0.388$            \\
            & d$Y$ & $-0.51\pm0.01$ & $0.02\pm0.05$  & $-0.81\pm0.02$ & $-0.42\pm0.01$ & $-0.24\pm0.01$ & $-0.01\pm0.01$ & $0.598$           & $0.496$            \\
   \hline
   changing & d$X$ & $-0.15\pm0.03$ & $1.56\pm0.21$  & $-0.04\pm0.04$ & $-0.18\pm0.02$ & $0.26\pm0.03$  & $-0.05\pm0.02$ & $0.658$           & $0.612$            \\
            & d$Y$ & $0.04\pm0.04$  & $-0.23\pm0.21$ & $-0.10\pm0.05$ & $-0.12\pm0.02$ & $0.01\pm0.02$  & $-0.44\pm0.03$ & $0.782$           & $0.703$            \\
   \hline
  \end{tabular}}
\end{table*}
The fitting results using model 3 are presented in Table~\ref{LLR-window_test}. The sliding-window method reduces the WRMS of the series, and the estimated deviations of all coefficients are smaller than those in Table~\ref{tab_LLR}. The changing-window method results in a lower accuracy in the CPO, \textcolor{Black}{but not in the fit coefficients. Moreover, the weighted means of the uncertainties of the two CPO series are 0.051 mas for the sliding-window method and 0.112 mas for the changing-window method, respectively. These results are within expectations. Observations are both more frequent and more accurate after 1985, so that with the sliding-window method we obtain a larger portion of the CPO of smaller uncertainties and thus reduce the weighted average. As for the changing-window method,} the window durations are different while many parameters change with time in the calculation process. According to these two tests, the limited quantity of obtained CPO corrections is a cause of the estimate uncertainties of the nutation terms. Moreover, the time distribution of the observations and the uncertainties of the models that were used in the calculation process are clearly not perfect.
\subsection{\textcolor{Black}{Future improvements of LLR observations for a better determination of the CPO}}
\begin{figure}[h]
 \label{sig_n}
 \centering
 \includegraphics[width=8cm]{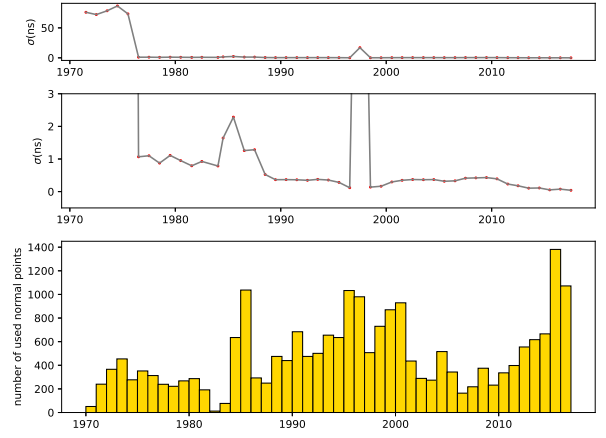}
 \caption{\textcolor{Black}{Time series of observational errors and number of used normal points; the second figure shows details of the first one. Each point \textcolor{Black}{or} bin represents the weighted average of the total within a year.}}
 \label{fig_sig_n}
\end{figure}

\textcolor{Black}{The accuracy of the CPO determined from LLR observations can be affected by many aspects. Of these, the observational error and frequency are most directly related to the observation itself. We show their changes with time in Fig.~\ref{fig_sig_n}. The observational accuracy, although it suffers from instability, is improving. By comparing Figs.~\ref{fig_sig_n} and~\ref{CPO_LLR_70d}, especially between 2000 and 2010, we can conclude that the dispersion and \textcolor{Black}{larger} uncertainties of the CPO in this period are the consequences of the lower frequency of the observations. Therefore, making LLR observations regular and sufficiently frequent to achieve a more uniform time distribution of normal points is quite essential in the future, before other necessary developments of related theories are possible.}
\section{Concluding remarks}
We presented comparative studies using VLBI, LLR, and optical data to estimate a possible improvement of the current precession-nutation models in terms of long-period nutation. VLBI, LLR, and optical data were first treated separately \textcolor{Black}{and then combined (LLR and VLBI only)} for further discussion. The \textcolor{Black}{weighted average of the standard error in the LLR series is comparable with} that of VLBI series, and the number rises to over \textcolor{Black}{two hundred} in the optical series.

The VLBI data present a \textcolor{Black}{most reliable} feature of the CPO series, and are still the most important data for determining the precession-nutation of Earth, \textcolor{Black}{whereas the formal errors may be underestimated}. The fit coefficients of the secular terms with the opa2018a series show an underestimation of the precession rate in d$X$ by about $0.3$ mas \textcolor{Black}{cy$^{-1}$}. All fit coefficients of the 18.6-year nutation term \textcolor{Black}{revealed an underestimate of} the amplitude by about $35$ \si{\mu as}.

The correlation between 18.6-year and 9.3-year nutation terms in LLR fitting results is very strong \textcolor{Black}{because of the direct relation between LLR observations and the motions of the Moon}. The fit coefficients show no evident correction to the precession rate, but a deviation of about $-0.3$ mas for all of the coefficients of the 18.6-year nutation term.

\textcolor{Black}{The differences between the details of the combined series and the smoothed VLBI series indicate that LLR has a comparable accuracy with VLBI in its best state.} The differences of the fit coefficients are almost the same as the deviations.

\textcolor{Black}{The potential of LLR for determining the CPO, and therefore the long-period part of precession-nutation, is confirmed, while the instability of the observational accuracy and the irregularity of observations, in particular, limit the consistency of the accuracies\textcolor{Black}{, as well as the time resolution} of the derived CPO series. Furthermore, the non-negligible correlations between the fit terms remind us that LLR is not yet dependable enough to contribute significantly to the improvement of the current precession-nutation model.}

In terms of the optical series, different from the results of VLBI series, the 18.6-year nutation term is not statistically significant. Considering the poor accuracy of optical observations, \textcolor{Black}{it is not reasonable to use optical observations to evaluate the contemporary precession-nutation models}.

\textcolor{Black}{In short, corrections to nutation terms in current models are suggested by VLBI observations, whereas the formal errors may be underestimated. Meanwhile, the LLR technique can be a candidate to provide an independent check for the determination of EOP after major improvements in observation regularity. Historical optical data are no longer helpful in spite of their long history.}

\begin{acknowledgements}
 \textcolor{Black}{The 13th version of software routines from the IAU SOFA Collection~\citep{Hohenkerk2012}\footnote{\url{http://www.iausofa.org}} was used to calculate the precession-nutation matrix according to the IAU\,1976/1980 and IAU2006/2000A models, and to convert  timescales as well.} We are grateful to D. A. Pavlov and his colleagues for their help and valuable advice in analyzing LLR data, and Niu Liu and Ping-Jie Ding for their suggestions to improve the manuscript. This research is funded by the National Natural Science Foundation of China (NSFC) No. 11473013 and No. 11833004. \textcolor{Black}{We also appreciate the detailed and valuable advice of the anonymous referee to improve this work.}
\end{acknowledgements}
\bibliographystyle{aa}
\bibliography{CPOreference}
\begin{appendices}
\section{\textcolor{Black}{Correlation coefficients}}
Correlation coefficients of all least-squares fits are listed for reference.
\begin{table}[h]
 \centering
 \caption{Correlation coefficients of the VLBI fitting results in Sect.~\ref{section_VLBI}.}
 \scalebox{0.9}{
  \begin{tabular}{rrrrrr}
   \hline\hline
   Term        & $t^0$ & $t^1$ & sin(18.6yr) & cos(18.6yr) & sin(9.3yr) \\
   \hline
   $t^1$       & -0.3  &       &             &             &            \\
   $t^2$       & -0.1  & -0.9  &             &             &            \\
   \hline
   $t^1$       & -0.8  &       &             &             &            \\
   sin(18.6yr) & 0.0   & 0.3   &             &             &            \\
   cos(18.6yr) & 0.2   & -0.3  & -0.0        &             &            \\
   \hline
   $t^1$       & -0.8  &       &             &             &            \\
   sin(18.6yr) & -0.0  & 0.4   &             &             &            \\
   cos(18.6yr) & 0.1   & -0.2  & 0.1         &             &            \\
   sin(9.3yr)  & -0.0  & 0.2   & 0.1         & 0.3         &            \\
   cos(9.3yr)  & 0.1   & -0.2  & -0.3        & -0.2        & -0.0       \\
   \hline
  \end{tabular}}
\end{table}

\begin{table}[h]
 \centering
 \caption{Correlation coefficients of the LLR fitting results in Sect.~\ref{section_LLR}.}
 \scalebox{0.9}{
  \begin{tabular}{rrrrrr}
   \hline\hline
   Term        & $t^0$ & $t^1$ & sin(18.6yr) & cos(18.6yr) & sin(9.3yr) \\
   \hline
   $t^1$       & -0.9  &       &             &             &            \\
   sin(18.6yr) & 0.1   & 0.3   &             &             &            \\
   cos(18.6yr) & 0.4   & -0.4  & 0.4         &             &            \\
   \hline
   $t^1$       & -0.4  &       &             &             &            \\
   sin(18.6yr) & 0.6   & 0.4   &             &             &            \\
   cos(18.6yr) & 0.7   & 0.0   & 0.7         &             &            \\
   sin(9.3yr)  & 0.6   & 0.3   & 0.7         & 0.8         &            \\
   cos(9.3yr)  & -0.6  & -0.2  & -0.8        & -0.5        & -0.5       \\
   \hline
  \end{tabular}}
\end{table}

\begin{table}[h]
 \centering
 \caption{Correlation coefficients of the optical fitting results in Sect.~\ref{section_optical}.}
 \scalebox{0.9}{
  \begin{tabular}{rrrr}
   \hline\hline
   Term        & $t^0$ & $t^1$ & sin(18.6yr) \\
   \hline
   $t^1$       & -0.6  &       &             \\
   $t^2$       & -0.8  & 0.3   &             \\
   \hline
   $t^1$       & -0.5  &       &             \\
   sin(18.6yr) & -0.1  & 0.1   &             \\
   cos(18.6yr) & 0.0   & 0.1   & -0.0        \\
   \hline
  \end{tabular}}
\end{table}

\begin{table}[h]
 \centering
 \caption{Correlation coefficients of the fitting results of the combined series in Sect.~\ref{section_joint}.}
 \scalebox{0.9}{
  \begin{tabular}{rrrrrr}
   \hline\hline
   Term        & $t^0$ & $t^1$ & sin(18.6yr) & cos(18.6yr) & sin(9.3yr) \\
   \hline
   $t^1$       & -0.2  &       &             &             &            \\
   $t^2$       & -0.2  & -0.9  &             &             &            \\
   \hline
   $t^1$       & -0.8  &       &             &             &            \\
   sin(18.6yr) & -0.1  & 0.4   &             &             &            \\
   cos(18.6yr) & 0.1   & -0.1  & 0.1         &             &            \\
   \hline
   $t^1$       & -0.7  &       &             &             &            \\
   sin(18.6yr) & 0.1   & 0.4   &             &             &            \\
   cos(18.6yr) & 0.0   & 0.0   & 0.1         &             &            \\
   sin(9.3yr)  & -0.1  & 0.2   & 0.1         & 0.3         &            \\
   cos(9.3yr)  & 0.1   & -0.0  & -0.2        & 0.0         & 0.1        \\
   \hline
  \end{tabular}}
\end{table}

\begin{table}[h]
 \centering
 \caption{Correlation coefficients of the LLR fitting results in Sect.~\ref{section_discuss}, of the sliding-window \textcolor{Black}{(the first part)} and changing-window \textcolor{Black}{(the second part)} methods.}
 \scalebox{0.9}{
  \begin{tabular}{rrrrrr}
   \hline\hline
   Term        & $t^0$ & $t^1$ & sin(18.6yr) & cos(18.6yr) & sin(9.3yr) \\
   \hline
   $t^1$       & 0.3   &       &             &             &            \\
   sin(18.6yr) & 0.7   & 0.3   &             &             &            \\
   cos(18.6yr) & 0.6   & 0.0   & 0.7         &             &            \\
   sin(9.3yr)  & 0.5   & 0.3   & 0.5         & 0.7         &            \\
   cos(9.3yr)  & -0.7  & -0.2  & -0.9        & -0.7        & -0.4       \\
   \hline\hline
   $t^1$       & -0.5  &       &             &             &            \\
   sin(18.6yr) & 0.6   & 0.4   &             &             &            \\
   cos(18.6yr) & 0.7   & -0.0  & 0.7         &             &            \\
   sin(9.3yr)  & 0.5   & 0.4   & 0.6         & 0.7         &            \\
   cos(9.3yr)  & -0.5  & -0.2  & -0.8        & -0.5        & -0.5       \\
   \hline
  \end{tabular}}
\end{table}
\end{appendices}

\end{document}